\documentclass[10pt,conference]{IEEEtran} 

\makeatletter
\def\endthebibliography{%
	\def\@noitemerr{\@latex@warning{Empty 'thebibliography' environment}}%
	\endlist
}

\IEEEoverridecommandlockouts
\usepackage{cite}
\usepackage{amsmath,amssymb,amsfonts}
\usepackage{algorithmic}
\usepackage{graphicx}
\usepackage{epstopdf} 
\usepackage{epsfig}
\usepackage{textcomp}
\usepackage{xcolor}
\usepackage{subfigure}
\def\BibTeX{{\rm B\kern-.05em{\sc i\kern-.025em b}\kern-.08em
    T\kern-.1667em\lower.7ex\hbox{E}\kern-.125emX}}
\definecolor{airforceblue}{rgb}{0.36, 0.54, 0.66}

\begin{document}
\title{Self-Supervised WiFi-Based Activity Recognition}
\author{\IEEEauthorblockN{Hok-Shing Lau, Ryan McConville, Mohammud J. Bocus, Robert J. Piechocki and Raul Santos-Rodriguez }	
	\IEEEauthorblockA{School of Computer Science, Electrical and Electronic Engineering, and Engineering Maths,  University of Bristol, UK.}
	\{sk19152, ryan.mcconville, junaid.bocus, eerjp, enrsr \}@bristol.ac.uk.
}

\date{\today} 

\maketitle



\begin{abstract}
Traditional approaches to activity recognition involve the use of wearable sensors or cameras in order to recognise human activities. 
In this work, we extract fine-grained 
physical layer information from WiFi devices
for the purpose of passive activity recognition in indoor environments. While such data is ubiquitous, few approaches are designed to utilise large amounts of  unlabelled WiFi data. We propose the use of self-supervised contrastive learning to improve activity recognition performance when using multiple views of the transmitted WiFi signal captured by different synchronised receivers.
We conduct experiments where the transmitters and receivers are arranged in different physical layouts so as to cover both Line-of-Sight (LoS) and non LoS (NLoS) conditions.
We compare the proposed contrastive learning system with non-contrastive systems and observe a 17.7\% increase in macro averaged $F_1$ score on the task of WiFi based activity recognition, as well as significant improvements in one- and few-shot learning scenarios.

\end{abstract}


\section{Introduction}
With the emergence of the Internet of Things (IoT), the development of contextual human sensing applications has become increasingly popular. Research in this area has made significant progress recently. For example, applications such as health monitoring within indoor environments based on Human Activity Recognition (HAR) has become feasible \cite{vesta}.
\begin{figure}
	\begin{center}
		\includegraphics[width=0.5 \textwidth]{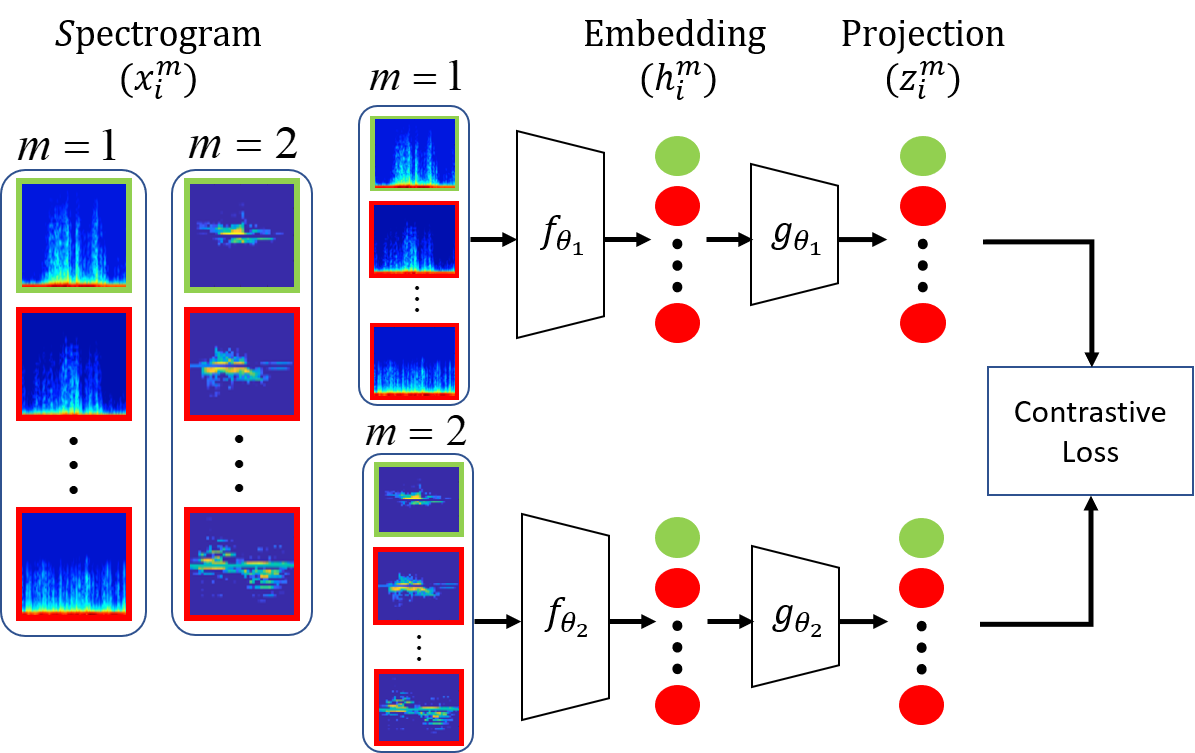}
    \caption{An overview of the WiFi based contrastive pretraining. In a feed-forward pass with a mini-batch of $N$ spectrogram samples, each sample $x_i$, $i\in\{1,2...N\} $ contains two views $m\in \{1,2\}$. In the figure, the green colours refer to a positive pair of samples, i.e., the signal at two different synchronised receivers, while the red refers to negative samples belonging to a different time-point. Each view of the sample $x^m_i$ is input into the corresponding encoder network $f_{\theta_{m}}$ to obtain the embedding $h^m_{i}$, which then further passes through the projection network $g_{\theta_{m}}$ to obtain the projection  $z^m_{i}$.
    After pretraining, the projection networks are discarded and the weights of the encoder networks frozen. For activity recognition a classification network is added on top of the encoders and fine-tuned with examples of labelled activities. 
    }
		\label{fig:pretraining}
	\end{center}
\end{figure}
Several WiFi sensing techniques have the potential for achieving non-restrictive, privacy-friendly indoor activity sensing when compared to current technologies such as cameras or wearable sensors \cite{accel}. Among the various WiFi sensing techniques, WiFi Channel State Information (CSI) has gained  popularity in recent years because of its extensive coverage in modern indoor settings, while providing fine-grained physical layer information that is useful for activity recognition \cite{survey}. The idea behind radar sensing involves 
detecting and identifying physical layer changes (e.g., Doppler and phase shifts, multipath propagation, signal attenuation, etc.) in the radar signal. 

Recently, research in radio based human sensing has moved towards deep learning \cite{deepl_survey} approaches
which have demonstrated  success due their ability to learn in complex environments.
One of the most popular deep learning architectures used for radar and WiFi classification is the Convolutional Neural Network (CNN), which involves transforming the raw signal into an image-like format such as spectrogram. 

CNNs have shown much success in radar and WiFi sensing applications, such as activity recognition \cite{sample_CNN_1,sample_CNN_2,uobwificsi}, localisation \cite{Li2020}, and gesture recognition \cite{sample_CNN_4}. 
However, these approaches typically require significant amounts of training data, which can be costly to obtain.
Further, generalisation to different environments has proved to be particularly challenging for radar and WiFi classification \cite{EI} as the multipath signal propagation is environment dependent. 
In this work, we propose to use a form of self-supervision, contrastive learning \cite{simclr}, in order to leverage synchronised data that is collected simultaneously. We investigate the potential for self-supervised contrastive learning to improve WiFi-based activity recognition performance, as well as the ability to improve performance with limited amounts of labelled data by leveraging synchronised WiFi data from multiple receivers.

The main contributions of  this work are the following:
\begin{itemize}

    \item We show that the proposed constrastively trained system can increase the activity recognition performance using a suitable network model architecture over a non-contrastive baseline.
    \item We assess the effectiveness of the proposed self-supervised system with limited labelled examples of each activity demonstrating that high performance can be achieved with relatively few labelled data points in a few-shot activity recognition scenario.

    \item We evaluate the effect of 
    the type of modality pairing on the contrastive learning performance, demonstrating that a constrastive pair of CSI outperforms a constrastive pair of CSI and Passive WiFi Radar (PWR).
\end{itemize}

The rest of this paper is organised as follows: 
Section II provides an overview of related work.
Section III 
describes the contrastive learning framework which is used as a pretraining process for HAR.
The experimental setup and system parameters are described in Section IV.
In Section V, we conduct several experiments to evaluate the effect of the pretraining process on the activity recognition performance ($F_1$ score) under different conditions, such as using different encoder networks, sampling in the fine-tuning phase, 
and modality pairing. 
Finally conclusions are drawn at the end of this paper.

\section{Related work}
In our previous work \cite{Sphere_1}, we performed a comprehensive
comparative study on the similarities and differences between
two WiFi sensing systems, namely, CSI and Passive WiFi Radar (PWR).
More specifically, we performed a direct comparison between CSI and PWR by concurrently capturing signals reflected off the bodies of five participants while they performed six different activities, namely, sitting, standing from chair, laying down, standing from lie and picking up an object from the floor. For a fair comparison, the raw signals from both the PWR and CSI systems were converted into spectrograms using signal processing techniques. 
A simple supervised 2D CNN was used as the classifier consisting of one convolutional layer, one max-pooling layer and two fully connected layers.
By considering data from three different physical layouts, consisting of a mixture of forward scatter (LoS), bistatic and monostatic layouts (NLoS), the CSI system achieved an overall accuracy of 67.3\% while the PWR system had a similar
accuracy at 66.7\%.
It was also shown that the CSI system had a better performance in LoS configurations (maximum accuracy of 90\%), whereas the PWR system had better performance in bistatic configurations (maximum accuracy of 91.3\%).

A common problem in supervised 
deep learning is the need for 
large amount of labelled data to train the 
parameters in the model.
Self-supervised learning \cite{simclr} is an effective strategy that leverages a large amount of unlabelled data to train the network without the use of labels.
Autoencoder-based pretraining is a popular approach in unsupervised learning that trains the encoder-decoder network to reconstruct a sample from its compressed form. This process is known as unsupervised pretraining, as the network learns a relevant representation from the data without the labels. To use this network for classification, the decoder of the network can be replaced with a classifier and trained under supervision, which is known as fine-tuning. 
With respect to WiFi and radar sensing, a number of works have shown that an autoencoder can be used for pretraining a deep neural network in applications such as localisation \cite{ConvAE}, user authentication \cite{Shi2017} and activity recognition \cite{Gao2017}.

\section{Representation learning with contrastive losses}

The goal of contrastive learning is to build a better data representation via judicious design of auxiliary tasks (here contrastive losses). Crucially, the entire pipeline is completely automated (`no labels'). In essence, we are attempting to build the function $ \bf{h}$: ${X \rightarrow Y}$, which maps the input data $X$ to some latent compact representation space $Y$. The mapping should maximise the mutual information $I(X;Y)$, while attempting to minimise the size of $\left| Y \right|$. As a result, $Y$ is made to capture the salient information about the data, while removing all spurious redundancy, so that subsequent classification tasks can be achieved relatively easily with just a handful of labelled examples (this is the the fine-tuning stage described below). However, direct maximisation of mutual information is a computationally intractable problem, and instead \cite{oord2019CPC} shows that lower bound on MI can be maximised by minimizing the contrastive loss i.e. $ -\mathcal{L}^{\text {cont}} < I\left(X; Y\right) $.   

The goal of contrastive loss $\mathcal{L}^{\text {cont}}$ is to minimise the difference between the learned representations of positive pairs of data and maximise the difference between the negative pairs \cite{simclr}. Generally, the positive pair consists of two samples of the same data point, which differ in some way, while the negative pair consists of two samples belonging to a different data point.
For example, SimCLR \cite{simclr} is a popular constrastive learning approach, originally proposed in computer vision under a self-supervised setting  and has been shown to be very successful.
Under the paradigm, a batch of images undergoes augmentation, such as rotation, crop and colouration. During the training, the two augmented samples originating from the same image are described as the positive pair, while the two augmented samples originating from different images are identified as the negative pair.
Contrastive learning has been used to learn crossmodal representations of audio-visual information \cite{cl_auto_visual} and to learn spatio-temporal features of the scenes \cite{cl_timecl}. 

In terms of radar sensing, \cite{cl_Federated} introduced scalogram signal correspondence learning. A scalogram is generated from a raw signal, and the network learns to align it with the corresponding raw signal with contrastive objective. 
In the application of WiFi-CSI, the authors demonstrate its ability to improve the generalisation under semi-supervised setting, where a small amount of labelled data is used to fine-tune the pretrained network, however the improvement is not substantial. 
\cite{cl_sense} examines eight strategies of contrastive learning in different activity recognition datasets. In the application of WiFi-CSI, the authors demonstrate that the networks trained with contrastive objectives have a competitive performance when compared to autoencoders.

\section{Contrastive WiFi-Based Activity Recognition System}

For our contrastive WiFi activity recognition we propose to use the synchronised data that is collected simultaneously from different receivers, or views, as the pairs. We note that the synchronised data is also collected from different angles and with different modalities.  
The training objective is to encourage the representations of the synchronised data, the different views, to be closer in the embedded space. These views represent the identical semantics of the environment and the activity at a given time-point. We propose that by training the model constrastively in this way, the model learns the inherent features that are invariant to the noise involved in such systems. Further, we propose that this form of self-supervision will better utilise the data collected and improve activity recognition performance using WiFi-CSI.

For the CSI system, the raw physical layer CSI measurement is obtained from a commercial Network Interface Card (NIC) and stored for off-line processing. The CSI is the channel estimate which is used 
during the equalisation stage in the WiFi receiver to reverse the effects of the channel (e.g., multipath propagation, attenuation, phase shift, etc) on the transmitted signal. 
For a WiFi system with multiple transmitting and receiving antennas, the CSI is obtained as a matrix consisting of complex values for each Orthogonal Frequency Division Multiplexing (OFDM) subcarrier.
Conversely, the raw WiFi signal in the PWR system is measured using a USRP platform, and is down-converted and digitised for real-time processing in a computing device. 
The PWR system correlates the transmitted signal and received signal to detect the Doppler shift and propagation delay. PWR uses a `reference channel' to recover the transmitted signal, and several `surveillance channels' to capture the reflected signals from different angles to provide spatial information. 
The signal processing pipeline which is used to convert the raw CSI and PWR data into spectrograms is fully described in
our previous work \cite{Sphere_1}.
We combine SimCLR \cite{simclr} and the Contrastive Multiview Coding \cite{multiviewcoding} to form our  framework for the contrastive learning. 
Fig. \ref{fig:pretraining} shows the overview of the pretraining contrastive learning stage. 
With the multi-view spectograms as input, we use the Normalised Temperature Cross-entropy (NT-Xent) proposed by \cite{simclr} to calculate the contrastive loss, which works as follows. As we train with mini-batches, we obtain $2N$ projections that result from applying a projection network to our embeddings of each pair, from each mini-batch with $N$ samples with two views. We then form positive pairs if the two projections originate from the same time point but different views, and negative pairs otherwise. Under the mini-batch, each projection forms exactly one positive pair and $2N-1$ negative pairs. To calculate the loss, we first compute the pairwise similarity $s$ for every available pair of projections $z_i$ and $z_j$ as follows:
\begin{equation}
    s_{i,j} = \frac{z^T_i\cdot z_j}{||z_i||||z_j||}.    \label{eq:pairwise similarity}
\end{equation}
We apply an exponential function on each pair, and divide the positive pair by the sum of the negative pairs. Then we take the negative logarithm to obtain the cross entropy, which is expressed as:
\begin{equation}
    \mathcal{L}^{\text {cont}}_{i} = -\log \frac{\exp(s_{i,j}/\tau)}{\sum_{k=1[k\neq i]}^{2N} \exp(s_{i,j}/\tau)},
\end{equation}
where $\tau$ is the softmax temperature.
Once we obtain the contrastive losses, $\mathcal{L}^{\text {cont}}_{i}$, for each projection in the mini-batch, we take the mean value of the losses to backpropagate through the networks. After pretraining, we discard the projection heads, freeze the encoders, add a classification network on top of the encoders and fine-tune them with labelled activity samples.

\section{Experiments}

\begin{figure}[t]
	\begin{center}
		\includegraphics[width=0.45 \textwidth]{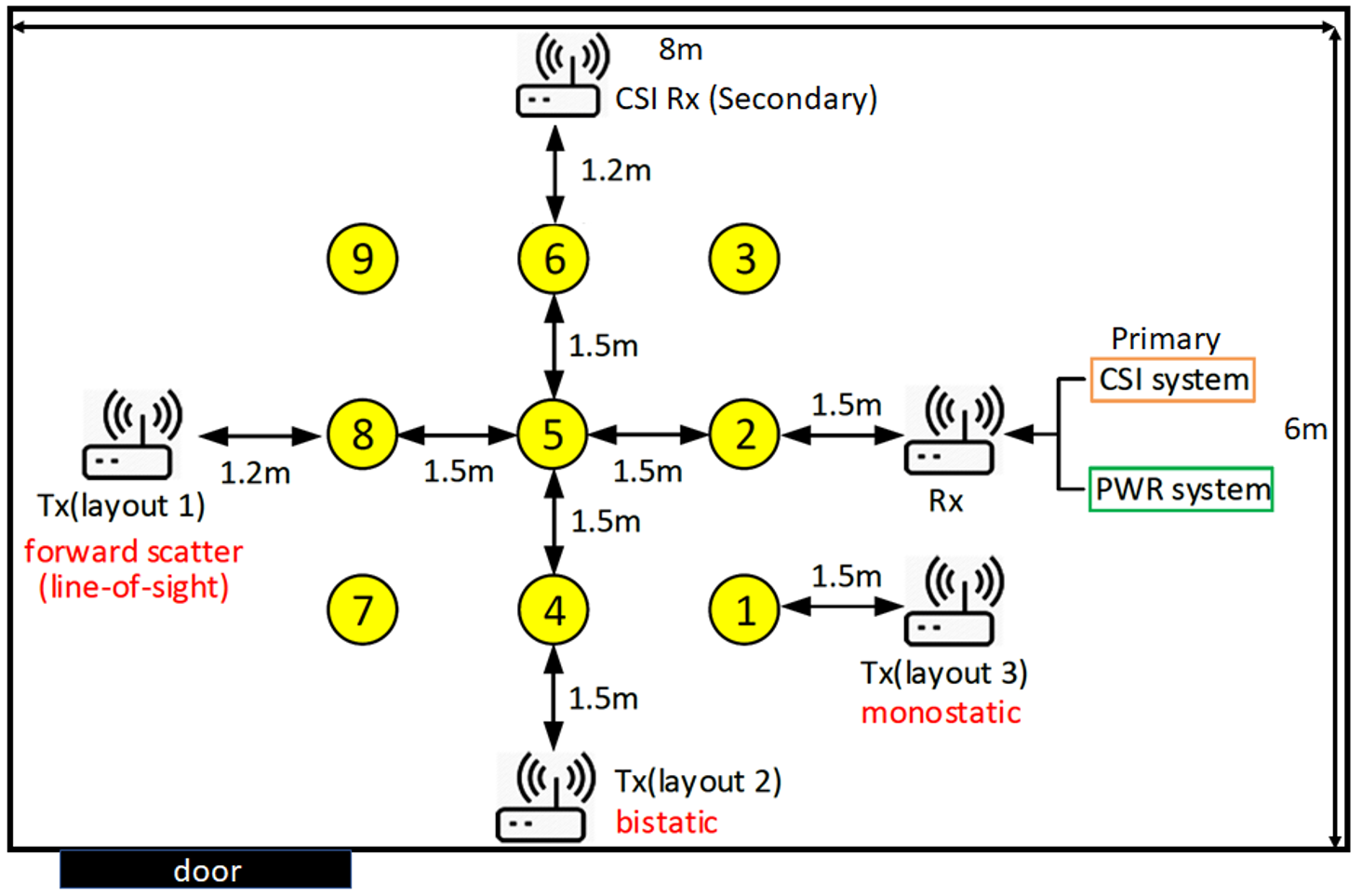}
    \caption{Layout used for the experiment. }
    \label{fig:exp_setup}
	\end{center}
\end{figure}

\noindent\indent\indent  
During the data collection, we used two WiFi CSI receivers (primary and secondary) and one PWR receiver to capture the WiFi signal emitted from a WiFi access point (transmitter). The working principles of the CSI and PWR systems are thoroughly described in \cite{Sphere_1}.
Locations of the receivers are varied based on the three layouts as depicted in Fig. \ref{fig:exp_setup}. 

Nine positions were tested during the experiment inside the monitoring area, with 1.5m separation between the
positions. 
The secondary WiFi CSI receiver was always located at 90${^0}$ to the primary CSI receiver in a bistatic geometry, while the 
PWR receiver and primary CSI receiver are collocated in each layout.
Both CSI receivers consist of an Intel 5300 Network Interface Card (NIC) \cite{csitool}, and the PWR receiver is implemented using a Software Defined-Radio (SDR) platform. 
Both systems operated in the 2.4 GHz WiFi band.

We collected the CSI and PWR data for seven different activities: laying down, picking something up from the ground (pickup), sitting (sit), standing from sit (stand), standing from the floor (standff), walking (walk) and waving, across five human subjects. 
Readers are referred to our previous work \cite{Sphere_1} for an in-depth description of the signal processing steps applied to the raw data from the CSI and PWR systems to obtain spectrograms. These serve as input to our contrastive learning system.
As we consider different layouts and positions (consisting of a mixture of forward scatter (LoS), bistatic and monostatic configurations (NLoS)) that affect the multipath effect considerably, this is more challenging than other studies which have only considered a single optimum layout.

\subsection{Contrastive Model Details}

Our system is implemented in PyTorch. During experiments we test three popular CNN architectures as the encoder: AlexNet \cite{AlexNet}, VGG16 \cite{VGG16} and ResNet18 \cite{ResNet} and one shallow network to analyse the effects of model architecture on the 
contrastive learning  performance.
The shallow network consists of three convolutional layers with 32, 64 and 96 filters, respectively. Each convolutional layer is followed by a batch normalisation layer and a max pooling layer. 
ReLU is used as the activation function for the first two layers while Tanh is used as the activation function for the third convolutional layer.
The dimensionality of the CSI and PWR spectrogram data is 65x501 and 100x41, respectively.
Because our input data sizes are comparatively smaller than typical image sizes 
used in the above-mentioned networks, 
we upscale our input spectrogram data by a factor of two or three using 2D nearest neighbour upsampling. 
Finally, the classifier is multi-layer perceptron with two hidden layers consisting of 128 and 7 neurons, respectively, while the projection head is a linear layer with a size of 128 neurons.

\subsection{Training and Evaluation}
We randomly select 80\% of the samples in the dataset as the training set and the remaining 20\% is used to evaluate the model.
All of the training set was used in the pretraining phase for contrastive learning. 
In addition to this, we also used subsets of the dataset with one, five and ten labelled  examples per class for fine-tuning to evaluate the few-shot learning capability of the model.
Due to the class imbalance in the HAR dataset, we use the macro averaged $F_1$ score for the majority of the evaluation.

\section{Results}


\begin{figure}[t]
	\begin{center}
		\includegraphics[width=0.40 \textwidth]{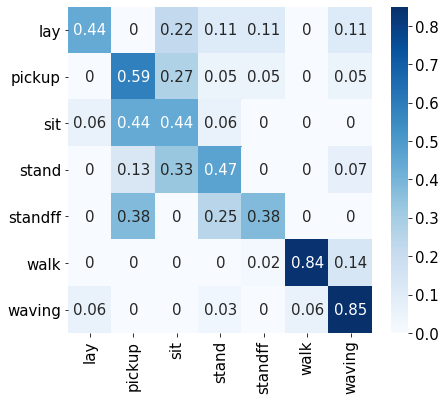}
    \caption{A confusion matrix demonstrating the performance of the non-contrastively trained system.
    }
		\label{fig:cmtx_normal}
	\end{center}
\end{figure}

\begin{figure}[t]
	\begin{center}
		\includegraphics[width=0.40 \textwidth]{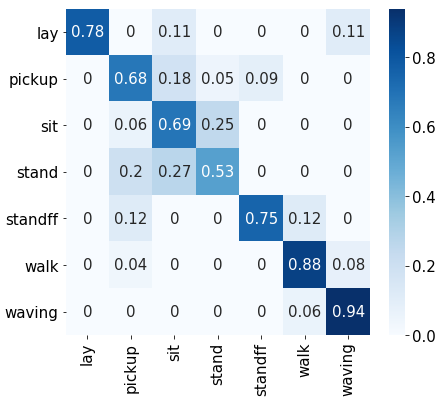}
    \caption{A confusion matrix demonstrating the performance of the contrastively-trained system. Improvements in performance can be seen across all activities.}
    \label{fig:cmtx_simclr}
	\end{center}
\end{figure}

\begin{figure}[t]
	\begin{center}
		\includegraphics[width=0.48 \textwidth]{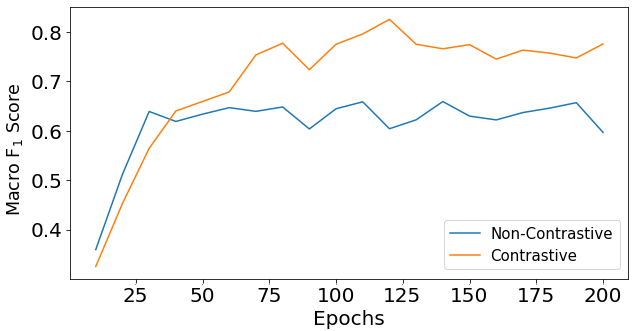}
    \caption{A comparison of the validation $F_1$ score (macro averaged) over 200 epochs between the non-contrastively-trained system and the contrastively-trained system.}
    \label{fig:performance}
	\end{center}
\end{figure}

We first present the results in 
Fig. \ref{fig:performance} which compares the macro averaged $F_1$ validation score of the proposed contrastive learning based model and the baseline, which is an AlexNet-based CNN without the contrastive pretraining step, over 200 epochs.
Both approaches are based on an AlexNet architecture, and consider as input WiFi CSI from two different receivers.
From this it is clear that the contrastively trained system outperforms a similar system without the contrastive pretraining step.

Figures \ref{fig:cmtx_normal} and \ref{fig:cmtx_simclr}  show the confusion matrices of the non-contrastively and contrastively pretrained systems, representing final $F_1$ scores of 57.9\% and 75.6\%, respectively.
In terms of the accuracy of the individual activity, the activity `walk' achieved the smallest improvement of 4.0\%, whereas the activity  `standff' obtains the largest improvement of 37.5\%, followed by `lay' (33.3\% increase) and `sit' (25.0\% increase).
Overall, the introduction of contrastive learning lead to a performance increase in the accuracy of 11.7\% and $F_1$ score of 17.7\%.

\begin{figure}
	\begin{center}
		\includegraphics[width=0.4 \textwidth]{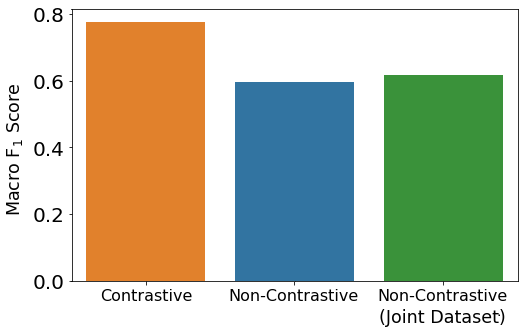}
    \caption{A comparison of the contrastively trained approach, in terms of macro $F_1$ score, with two versions of the non-contrastively trained approach. The `Non-Contrastive' approach uses only data from a single CSI receiver (CSI-1), while `Non-Contrastive (Joint Dataset)' uses data from both CSI receivers (CSI-1 and CSI-2) but is still non-contrastively trained.    
    }
    \label{fig:compare_method}
	\end{center}
\end{figure}

To further evaluate the effectiveness of our contrastive learning system, we compare it with a non-contrastively trained version under different training regimes, as illustrated in Fig.  \ref{fig:compare_method}.
In this experiment, `Non-Contrastive' refers to CSI data from the primary receiver (CSI-1) only while 
`Non-Constastive (Joint Dataset)' refers to CSI data from both receivers (CSI-1 and CSI-2) but is still not contrastively trained.
It can be observed from this figure that while the inclusion of the second CSI receiver slightly improves performance under a normal supervised training regime, contrastive pretraining leads to a more significant improvement in performance.

\subsection{Sample Efficiency}

\begin{figure}
	\begin{center}
		\includegraphics[width=0.5 \textwidth]{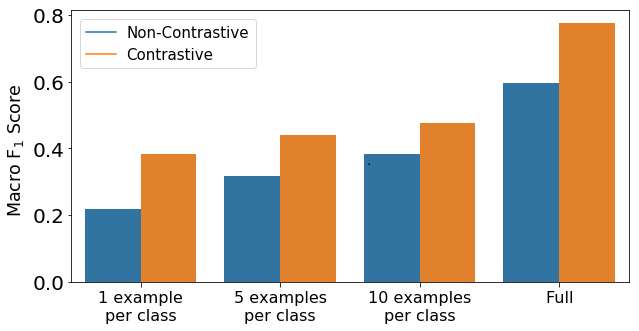}
    \caption{A comparison of the performance of the contrastively trained system and the non-contrastively trained system in scenarios with limited amounts of labelled data.
    }
    \label{fig:comapre_sampling}
	\end{center}
\end{figure}

\begin{figure}
	\begin{center}
		\includegraphics[width=0.5 \textwidth]{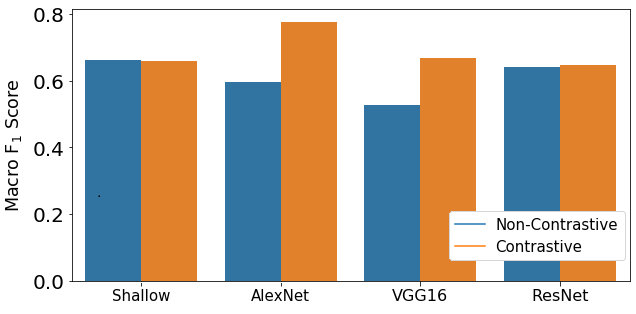}
    \caption{A comparison of the effect of contrastive pretraining using different encoder architectures. }
    \label{fig:compare_model}
	\end{center}
\end{figure}
Next, we evaluate the performance of the system in one-shot and few-shot activity recognition scenarios. The results of this can be seen in Fig. \ref{fig:comapre_sampling}, where the number of labelled samples in the classification stage was reduced to one, five and ten examples of each activity. 
The results demonstrate that the use of contrastive pretraining on synchronised WiFi CSI from multiple receivers can significantly improve activity recognition performance in one- and few-shot learning scenarios.

\subsection{Encoder Architectures}
Fig. \ref{fig:compare_model} compares the performance of the non-contrastive and contrastively pretrained models with different encoder architectures. 
From this, we can observe that the difference in $F_1$ score between non-pretrained and pretrained models is largest with an AlexNet-based CNN encoder.
Furthermore, the same encoder also achieves the best overall activity recognition performance.
On the other hand, little to no gain in performance can be observed with a `shallow' CNN based encoder, which has less total parameters than either the projection head and the classifier, as well as a ResNet based encoder.
Further investigation of this is left for future work.

\begin{figure}
	\begin{center}
		\includegraphics[width=0.4 \textwidth]{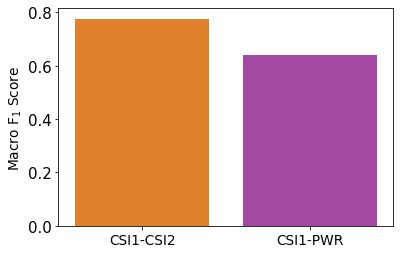}
    \caption{A comparison of the effect of the design of pairs on contrastive learning performance. The proposed system uses two pairs of CSI (CSI-1 and CSI-2), as it outperforms the same approach when constructing pairs from CSI and PWR.
    }
    \label{fig:compare_paring}
	\end{center}
\end{figure}

\subsection{Modalities}
Finally, we study the effect of the type of modality pairing on the contrastive learning performance. The proposed contrastively pretrained WiFI HAR system uses data from the same modality to construct pairs, that is, a positive pair consists of CSI data from the two synchronised CSI receivers, CSI-1 and CSI-2.
In our experiments, we also collected Passive WiFi Radar (PWR) data, as can be seen in Fig. \ref{fig:exp_setup}. Thus, we test the performance of the same contrastively pretrained system, but by constructing pairs of WiFI CSI (CSI-1) and PWR data.

There was a significant difference in the macro averaged $F_1$ score, as shown in Fig. \ref{fig:compare_paring}.
This experiment shows that contrastively training on synchronised CSI pairs outperforms training on CSI PWR pairs.

\section{Conclusions}
In this work, we propose a system which uses
contrastive learning
on WiFi data to improve activity recognition performance. Specifically, the system utilises multiple views from synchronised receivers for contrastive learning. 
We evaluate the performance of our proposed model using experimental data consisting of seven activities recorded from five human participants in different receiver layouts and positions. 
Through experimental evaluations we show significant improvement in the activity recognition performance using self-supervised contrastive learning when compared to conventional supervised models.
Specifically, contrastively pretraining with an AlexNet-based encoder lead to a 17.7\% increase in macro averaged $F_1$ score.
We also evaluated the proposed approach in one- and few-shot learning scenarios, observing contrastive learning with WiFi CSI pairs from synchronised receivers leads to a significant improvement in activity recognition performance.

\section*{Acknowledgements}
This work was performed as a part of the OPERA Project, funded by the UK Engineering and Physical Sciences Research Council (EPSRC), Grant EP/R018677/1. 


\bibliographystyle{IEEEtran}
\bibliography{ref}


\end{document}